\documentclass[12pt]{article}

\usepackage{newtxtext,newtxmath}
\usepackage{siunitx}
\usepackage{booktabs}
\usepackage{graphicx}
\usepackage[table,xcdraw]{xcolor}
\usepackage{adjustbox}
\usepackage[letterpaper,margin=1in]{geometry}
\usepackage{tabularx}
\usepackage{booktabs}
\usepackage{array}
\usepackage{adjustbox}
\usepackage{threeparttable}

\renewenvironment{abstract}
	{\quotation}
	{\endquotation}

\date{}

\makeatletter
\renewcommand{\fnum@figure}{\textbf{Figure \thefigure}}
\renewcommand{\fnum@table}{\textbf{Table \thetable}}
\makeatother

\usepackage{scicite}

\usepackage{url}

\def\scititle{
    Agonist–Antagonist Neural Coordination without Mechanical Coupling after Targeted Muscle Reinnervation
}
\title{\bfseries \boldmath \scititle}

\author{
	Laura~Ferrante$^{1\ast}$,\and
	Anna~Boesendorfer$^{2}$,\and
	Benedikt~Baumgartner$^{2}$,\and
    Manuel~Catalano$^{3}$,\and
    Antonio~Bicchi$^{3,4\dagger}$,\and
    Oskar~C.~Aszmann$^{2,5\dagger}$,\and
    Dario~Farina$^{1\ast\dagger}$\\
	\small$^{1}$Department of Bioengineering, Imperial College London, London, United Kingdom.\and
	\small$^{2}$Clinical Laboratory for Bionic Extremity Reconstruction, Department of Plastic, Reconstructive \\\small and Aesthetic Surgery, Medical University of Vienna, Vienna, Austria.\and
	\small$^{3}$SoftRobotics Lab for Human Cooperation and Rehabilitation, Istituto Italiano di Tecnologia, Genoa, Italy.\and
    \small$^{4}$Research Center E. Piaggio, University of Pisa, Pisa, Italy.\and
    \small$^{5}$Department of Plastic, Reconstructive and Aesthetic Surgery, Medical University of Vienna, Vienna, Austria.\and
	\small$^\ast$Corresponding authors. Email: l.ferrante@imperil.ac.uk;d.farina@imperial.ac.uk\and
	\small$^\dagger$Equal contribution for senior authorship.
}

\begin{document} 

\maketitle

\begin{abstract} \bfseries \boldmath

Following limb amputation and targeted muscle reinnervation (TMR), nerves that originally innervated agonist and antagonist muscles are rerouted into one or more residual target muscles. This rerouting profoundly alters the natural mechanical coupling and afferent signalling that normally link muscle groups in intact limbs. Despite this disruption, in this study we demonstrate, using high-density intramuscular microelectrode arrays implanted in reinnervated muscles of three TMR participants, that motor units (MUs) associated with agonist and antagonist tasks remain functionally coupled. Specifically, over 
40\% of motor units active during agonist tasks were also recruited during the corresponding antagonist tasks, even though no visual feedback on antagonist neural activity was provided. These motor units exhibited significantly different firing rates depending on their functional role. 
These results provide the first motor-unit–level evidence that the central nervous system preserves coordinated agonist–antagonist control after TMR and inform restorative surgical strategies and prosthetic systems capable of regulating both limb kinematics and dynamics based on agonist-antagonist commands interplay.
\end{abstract}

\section*{Introduction}
\noindent
Limb amputation leads to sensorimotor impairment, as both afferent and efferent neural pathways are irreversibly disrupted. Advanced nerve transfer techniques such as Targeted Muscle Reinnervation (TMR) \cite{kuiken2004use} and Regenerative Peripheral Nerve Interfaces (RPNIs) \cite{kung2014regenerative}, reconfigure the residual peripheral tissues to maximise the number of functionally relevant neural signals that can be harnessed for prosthesis control via surface or invasive electromyography (EMG).
Although TMR and RPNIs have a key role in reducing phantom limb pain and facilitating sensorimotor rehabilitation through artificial limbs, these procedures do not restore the neuromechanical coupling between groups of agonist and antagonist muscles typical of intact limbs.
\begin{figure}
\centering
\includegraphics[width=0.95\textwidth]{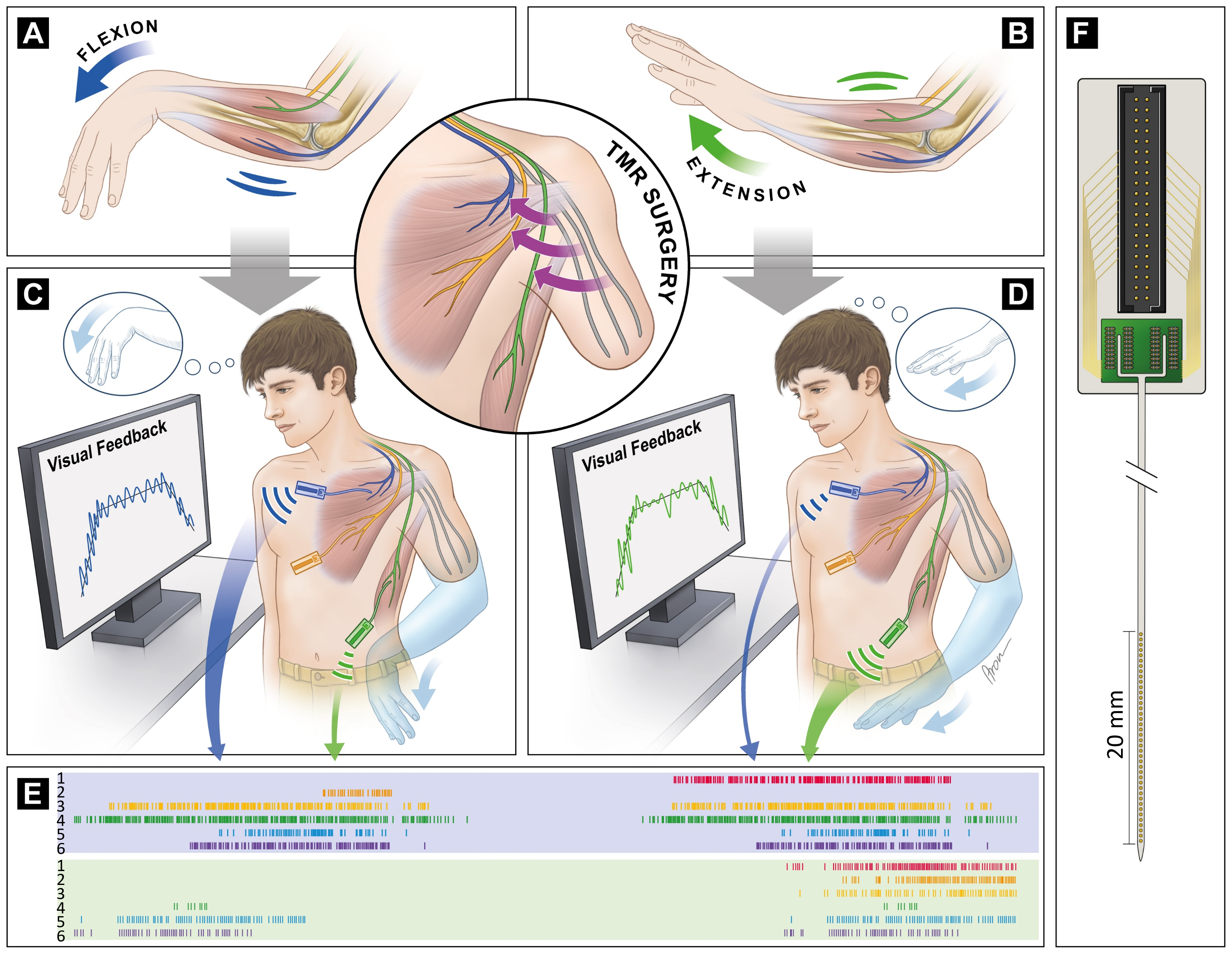}
\caption{\textbf{Microelectrode arrays record residual neural coupling in muscles reinnervated by transferred polyfascicular nerves during agonist and antagonist tasks.} 
\textbf{a-b}, In individuals with intact limbs, movement is controlled by coordinated activation of agonist and antagonist muscles spanning relevant joints. \textbf{c-d}, Amputation disrupts the neuro-mechanical coupling between agonist-antagonist muscles. P2 had the residual ulnar (blue), median (orange) and radial (green) nerves transferred to targeted muscles via TMR. P2 performed wrist flexion–extension while intramuscular activity of the agonist and antagonist (as defined for the task) reinenrvated myscles was recorded via percutaneous microelectrode arrays. Only the agonist reinnervated muscle was used as real-time visual feedback. P2 modulated the contraction of the agonist reinnervated muscle to match a trapezoidal activation profile. \textbf{e,} Neural signals recorded by the two microelectrode arrays (blue and green) during wrist flexion-extension are decomposed into motor unit spike trains. Motor units recruited when the reinnervated muscle acted as agonist are recruited when it acted as antagonist. \textbf{f,} Schematic of 40-channel high-density microelectrode array \cite{muceli2022blind}. Panels a–b show a representative agonist–antagonist pair for simplicity; neural structures and the microelectrode array are not to scale to improve clarity.}
\label{fig:concept}
\end{figure}
In TMR, severed nerves that carry agonist and antagonist signals are typically rerouted into biomechanically uncoupled targeted muscles (Fig.\ref{fig:concept}-a,d,c,d). Importantly, when viable, each nerve that once innervated multiple muscles of the missing limb and provided neural drives for different functional tasks (i.e., polifascicular nerve) is redirected into a single targeted muscle. In RPNI, nerves are surgically split into fascicles, each rerouted into a separate muscle graft. This nerve–muscle reorganisation eliminates the mechanical synergistic or antagonistic interactions between muscles that previously spanned the same articular joint, impairing afferent signalling mediated by spindles and Golgi tendon organs \cite{proske2012proprioceptive}. In turn, this hinders neural processes at the base of motor coordination, including reciprocal inhibition of motor neurons innervating agonist and antagonist muscles (although spindles reinnervation occurs in animal models of TMR \cite{festin2024creation}, its extent and functional role in humans remain unclear). As a result, amputees rely heavily on visual cues and incidental sensory inputs to compensate for impaired afferent feedback.

Most research on the coordination of residual agonist–antagonist muscle groups has focused on the lower limb, reporting highly variable patterns of activation and frequent coactivation between functionally opposing residual muscles \cite{huang2018voluntary}. In upper-limb amputees, EMG-based prosthesis studies show that control commands for agonist and antagonist movements can still be derived from their respective residual muscles (depending on residual peripheral tissue), a paradigm routinely used to control a single prosthesis function directionally. However, the functional significance of residual coactivation has been examined in a limited number of studies \cite{zuniga2018coactivation}. 
In individuals who have undergone nerve transfers, re-routed nerves, in principle, continue to convey neural signals associated with agonist and antagonist tasks to the targeted muscles or muscle grafts. In clinical practice, control of these opposing functions is typically achieved using EMG signals recorded from separate reinnervated muscles \cite{kuiken2004use}. While this demonstrates voluntary modulation of signals for opposing tasks, patterns of coactivation are rarely examined. Moreover, because conventional bipolar EMG electrodes are used to record from muscles reinnervated by polyfascicular nerves, the resulting signals likely reflect a mixture of neural commands, making it unclear whether the recorded neural activity underlying single-channel EMG signals truly preserves distinct functional roles such as agonist–antagonist specificity.
\begin{figure}
\centering
\includegraphics[width=1\textwidth]{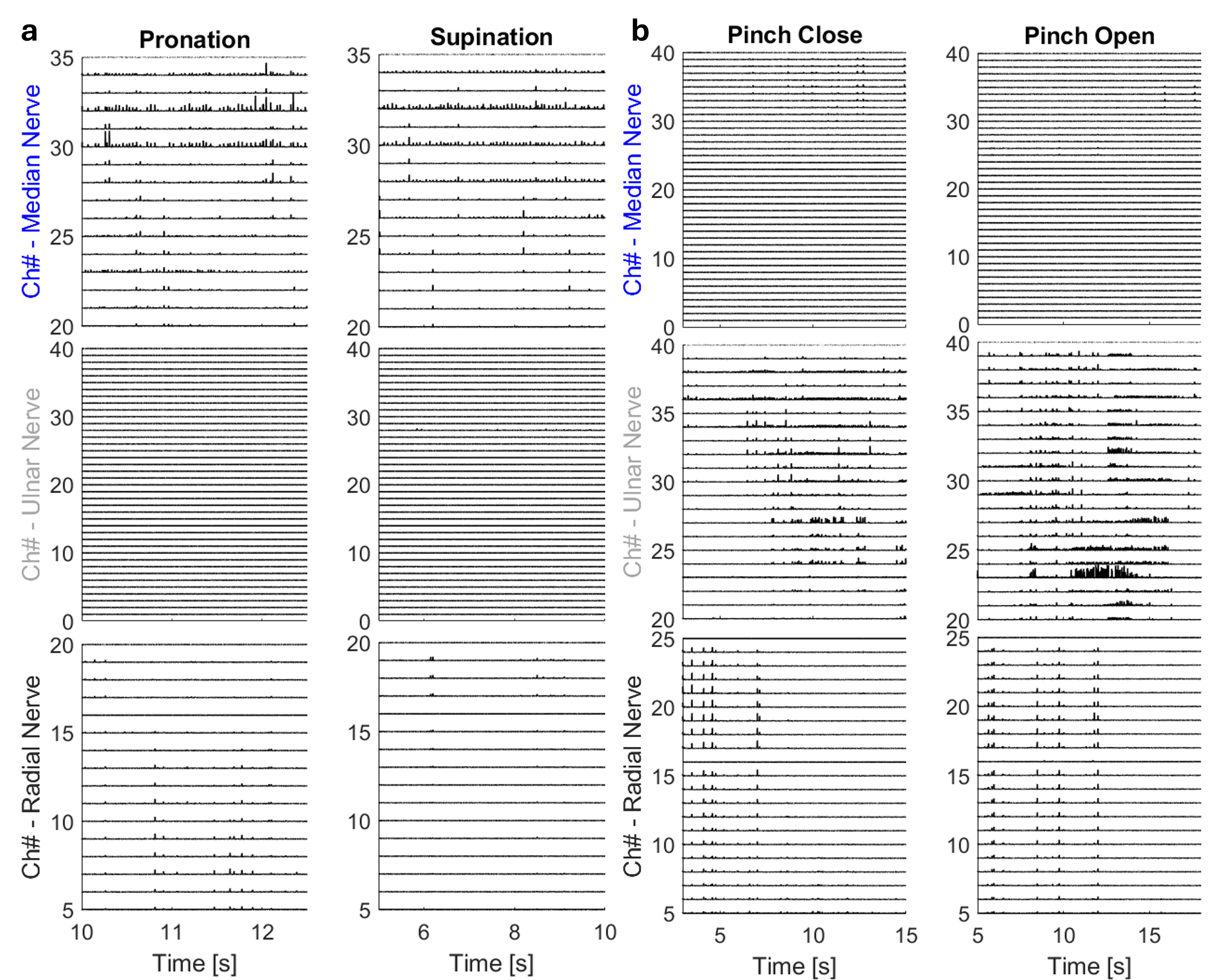}
\caption{\textbf{Intramuscular EMG recordings from three reinnervated muscles in P2.} The activity of three reinnervated muscles was concurrently recorded. The EMG envelopes obtained from each microelectrode array are shown for two pairs of tasks: pronation, supination (panel a); pinch open and pinch close (panel b). EMG amplidute is normalised per each pair of tasks considering the maximum EMG amplitude recorded across all microelectrode arrays. In some plots, time interval and channel range are chosen to best visualize the motor units spikes. In can be observed that for each task-pair, two reinnervated muscles are mainly active, while the third one either shows no activity or minimal recruitment depending on their synergistic contribution in the task. During pronation, the reinnervated muscle innervated by the median nerve is mainly active and therefore this muscle is classified as agonist for pronation. During supination the muscle reinnervate by the radial nerve is defined as antagonist.}
\label{fig:emgs}
\end{figure}

Here, we investigate the neural signals underlying co-activation in muscles reinnervated by polyfascicular nerves that formerly innervated agonist and antagonist muscle groups of the now-missing limbs, in the absence of any feedback on such co-activation. We hypothesized that the central nervous system continues to coordinate agonist and antagonist muscle commands. This would occur even when the nerves controlling these opposing actions were reinnervated into the same muscle, as a result of polyfunctional reinnervation. For example, the extensor carpi radialis brevis and the extensor carpi ulnaris are both supplied by the radial nerve and act as synergists for wrist extension, yet they behave as antagonists during hand abduction–adduction. We further hypothesised that such coordination would persist when commands were distributed across biomechanically separated muscles. For instance, the ulnar and radial nerves supply agonist–antagonist commands for wrist flexion and extension. If confirmed, such activation could be exploited not only for controlling advanced prostheses with variable stiffness actuators, allowing amputees to modulate the dynamics of movement besides the kinematics, but also for artificial substitution of sensory feedback associated with joint impedance and movement perception.
\section*{Results}
\subsection*{Motor unit recordings in muscles reinnervated by nerves carrying agonist and antagonist commands}

\begin{figure}
\centering
\includegraphics[width=0.9\textwidth]{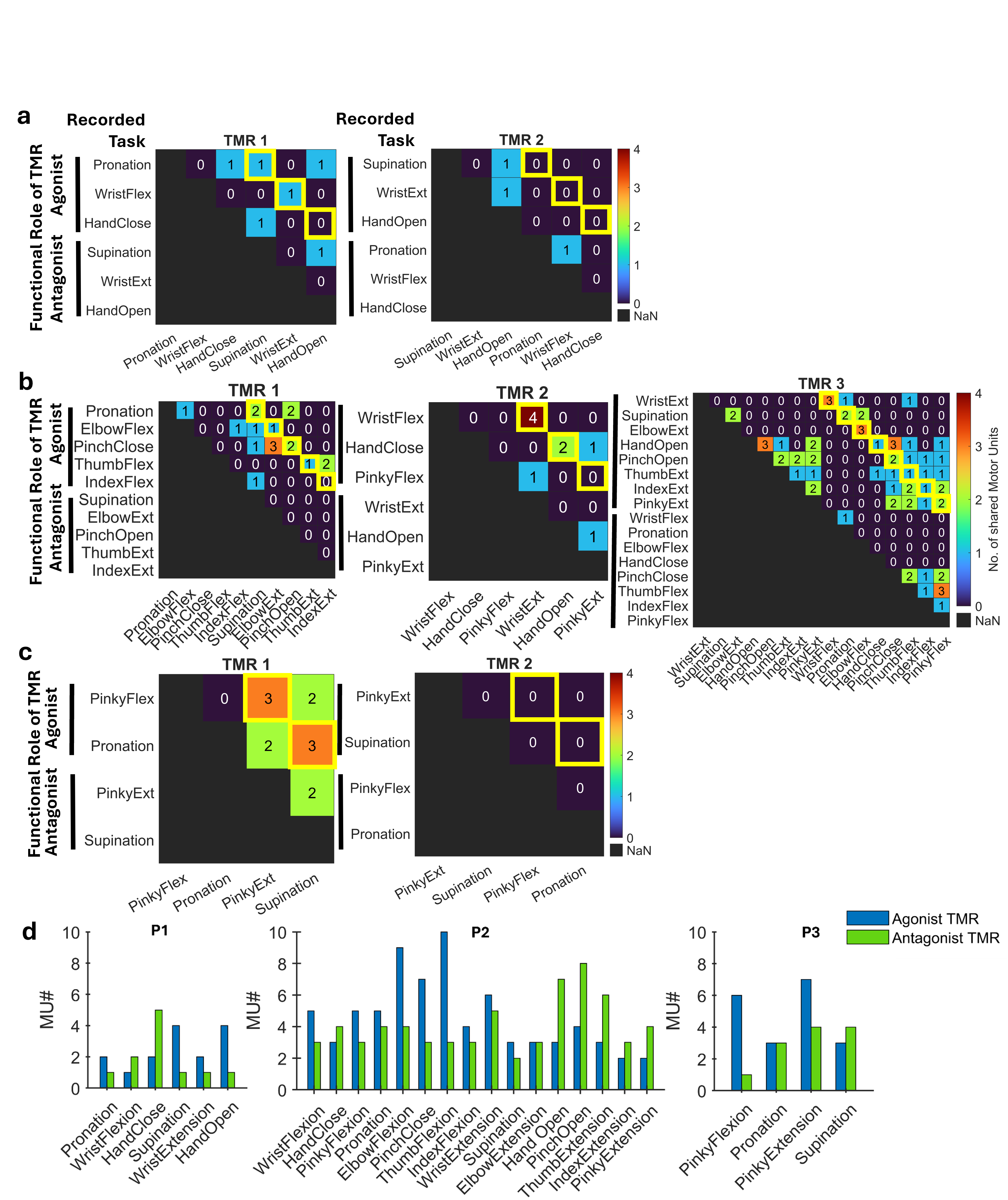}
\caption{\textbf{a,b,c,} For patients (P1, P2, P3) and reinnervated muscle (TMRX), the diagrams detail the relation between recorded tasks in terms of the number of shared motor units. Given a task, the agonist and antagonist TMR muscle was identified based on their rerouted nerve (e.g. when P2 performed wrist flexion, TMR2 acted as an agonist; TMR3 acted as agonist during wrist extension). Wrist extension was thus labelled as agonist in TMR3, but as antagonist in TMR1. Highlighted squares indicate the number of motor units active both when a TMR muscle functioned as an agonist and when the same muscle functioned as an antagonist during the opposite task. \textbf{d,} For each task, we report the total number of motor units identified for the agonist and antagonist reinnervated muscles.}
\label{fig:shared}
\end{figure}
The research hypotheses were investigated in three patients (P1, P2, and P3) who underwent polyfunctional reinnervation through TMR. Patient P3 (53 years old, glenohumeral amputation) had participated in a previous study \cite{ferrante2025implanted} in which we demonstrated a proof‑of‑concept biological interface combining polyfascicular nerve transfers via TMR with high‑resolution recordings from intramuscular microelectrode arrays. That study showed that distinct neural drives for several tasks of the missing-limb could be mathematically isolated using motor-unit decomposition and neural manifold analysis.
Building on this prior work, we re-analysed part of the data from patient P3, who had micro-electrode arrays inserted in the pectoralis minor (P3-TMR1) and in the latissimus dorsi muscles (P3-TMR2), reinnervated by the ulnar and radial nerves, respectively. Since only two pairs of agonist-antagonist tasks were recorded for P3, two additional participants (P1 and P2; aged 62 and 63, transhumeral amputation) were recruited specifically for the current study. In P1, the median and ulnar nerves were rerouted to the anterior deltoid (P1-TMR1), and the radial nerve to the posterior deltoid (P1-TMR2). In P2, the median nerve was transferred to the abdominal portion of the pectoralis major (P2-TMR1), the ulnar nerve to the pectoralis minor (P2-TMR2), and the radial nerve to the latissimus dorsi (P2-TMR3). This nerve–muscle reorganisation via TMR is illustrated for P2 in Figure~\ref{fig:concept}. As in the previous study, 40-channel microelectrode arrays \cite{muceli2010identifying} (Fig.\ref{fig:concept}-f) were inserted in each of the reinnervated muscles listed above. Additional details on amputation level, surgical procedures, and prosthesis use for each participant are provided in Table~\ref{tab:patient} (Section \textit{Materials and Methods}; refer to this same section for further details on the procedure used for microelectrode arrays insertion).

Biomechanically, during a specific movement at a given articular joint, flexor and extensor muscles spanning that joint can act as agonists or antagonists depending on the direction of the torque they generate relative to joint acceleration. Muscles producing torque in the same direction as the joint acceleration act as agonists, whereas those generating torque in the opposite direction act as antagonists. Within the agonist group, muscles may further be classified as prime movers or synergists according to the magnitude and functional relevance of their contribution. In this study, for each target movement of the missing limb that participants were asked to perform, we identified, among the reinnervated muscles, a single primary agonist and a single primary antagonist. This classification was based on the functional role of the corresponding nerve in the intact limb. For example, in P2 (Fig.~\ref{fig:concept}c,d), the muscle reinnervated by the ulnar nerve (blue) was designated as the primary agonist for wrist flexion and the primary antagonist for wrist extension, whereas the muscle reinnervated by the radial nerve (green) was designated as the primary agonist for wrist extension and the primary antagonist for wrist flexion. In P2, the definition of agonist and antagonist reinnervated muscles for each task is validated by inspecting the EMG signals recorded by all three microelectrode arrays. For the reminder of the paper, we define a reinnervated muscle as agonist or antagonist based on the above definition. In Fig.~\ref{fig:emgs}, we show EMG signals from two representative pairs of tasks (panels a,b). It can be observed that two reinnervated muscles are mainly active for each pair of tasks, while the third reinnervated muscle provides either no or little contribution, depending on its role within the task. This shows that the observed co-activations are not simply the effect of a generalized co-activation of all nerves irrespective of the task. Rather, the co-activation follows the expected agonist/antagonist roles of the originally innervated muscles.

The experimental paradigm was designed so that, when participants were instructed to perform a movement of the missing limb, they received visual feedback exclusively on the EMG activity of the reinnervated agonist muscle to eliminate voluntary modulation of antagonistic activity based on task (Fig.~\ref{fig:concept}c,d). Although visual feedback and task instructions were provided only for the agonist task, intramuscular signals were recorded concurrently from both the agonist and antagonist reinnervated muscles. This allowed us to characterize the intrinsic co-activation patterns, providing insight into the extent to which neural coupling between agonist and antagonist pathways was preserved and functionally expressed following targeted reinnervation and in the absence of mechanical coupling between agonist-antagonist muscles. 

P1 and P2 performed three and eight agonist–antagonist task pairs, respectively. For P3, we reused previously collected data comprising two task pairs. In total, this resulted in six, sixteen, and four recorded missing-limb tasks for P1, P2, and P3, respectively (see \textit{Materials and Methods}  for detailed experimental protocol).

For each microelectrode array, EMG channels yielding a baseline noise (root-mean-square of \SI{5}{\second} of data recorded at rest before starting the trials) greater than \SI{15}{\micro\volt} were visually inspected and removed. The retained channels across participants and task had an average baseline noise of  6.74 $\pm 1.62$ \SI{}{\micro\volt}. Intramuscular EMG signals were high-pass filtred with a zero-lag first-order digital filter at \SI{1000}{\hertz} and subsequently decomposed into individual motor unit spike trains through blind source separation \cite{muceli2022blind,farina2026decoding}. The decomposition outcome was refined by inspecting each spike train on each channel of all microelectrode arrays with EMGLAB spike-sorting software \cite{mcgill2005emglab}. Motor units with either few sparse firings or a low pulse-to-noise ratio were discarded (see \textit{Materials and Methods}).
\begin{figure}[!]
\centering
\includegraphics[width=0.9\textwidth]{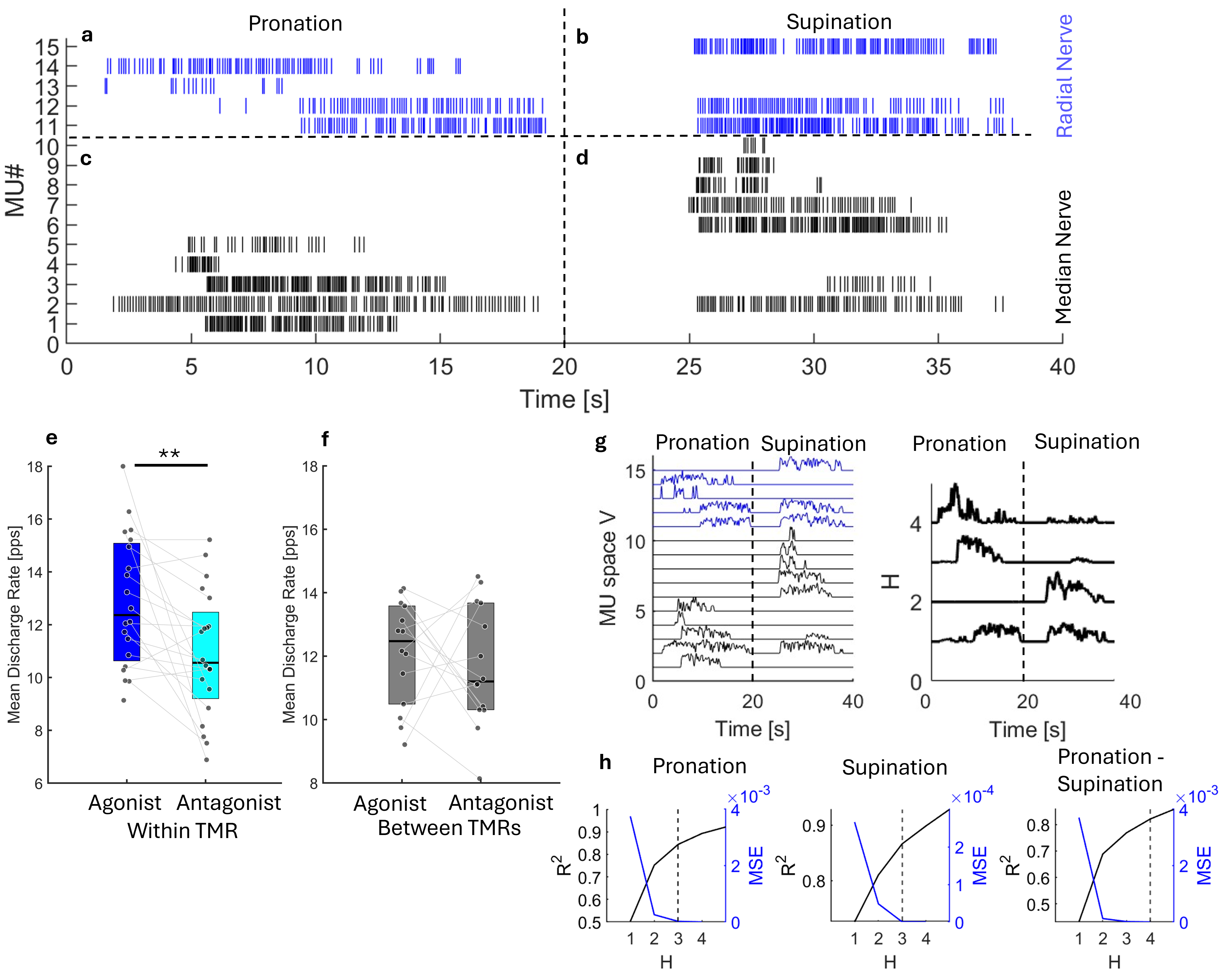}
\caption{\textbf{Discharge properties of MUs in agonist and antagonist reinnervated muscles of P2 and neural manifold analysis.} In \textbf{a-d}, MUs spike trains of muscle reinnervated by the median nerve (\textbf{a-b} - blue spikes) and radial nerve (\textbf{c-d} - black spikes). During pronation, agonist MUs are shown in (a) and antagonists in (c); during supination, in (d) and (b). Motor units recruited during both pronation and supination are tracked (e.g., in panel \textbf{a}, motor unit 11 is also active during supination in panel \textbf{b}). \textbf{e}, Mean discharge rate of each motor unit active in both a) and b) or c) and d). Each group contained 18 points (motor units from all tasks). The two-sided Wilcoxon signed-rank test is used to assess statistically significant differences between the two groups (p-value $<$ 0.005). Statistical significance is indicated with a bar and double asterisk.
\textbf{f,} Distribution of average mean discharge rate of motor units recorded in the agonist and corresponding antagonist reinnervated muscle for each task. 
\textbf{g,} On the left, a 15-dimensional MU spare (V) is obtained by concatenating smoothed discharge rates of motor units recruited during pronation and supination in TMR1 and TMR2. The dimensionality of the latent space (H) embedded in V is estimated using NNMF. The time-varying latent signals are shown on the left for the estimated dimensionality of 4. \textbf{h,} $R^{2}$-curve obtained by applying NNMF on $V$ with increasing latent space dimensionality. Dimensionality was 4 when pronation and supination data were concatenated, and lower when agonist and antagonist motor units were analyzed separately for each task.}
\label{fig:3}
\end{figure}
\subsection*{Motor units count and tracking across agonist-antagonist tasks}
For P1,2,3, considering all recorded tasks, a total of 15, 78 and 19 motor units were reliably decomposed from intramuscular recordings of agonist reinnervated muscles, respectively (average number of motor units per task 2.5$\pm$1.2\% in P1, P2 4.9$\pm$2.4\% and 4.5$\pm$2.4\%); 11, 61, 12 motor units were decomposed from corresponding antagonist reinnervated muscles (average number of motor units per task 1.8$\pm$1.6\% in P1, 3.8$\pm$1.5\% in P2, and 3.5$\pm$0.6\% in P3). Importantly, because participants received visual feedback exclusively on the EMG activity of the agonist reinnervated muscle, motor units detected in the antagonist reinnervated muscle reflect the intrinsic agonist–antagonist motor commands generated during task execution.

Motor units recruited during a given task and identified in either the agonist or antagonist reinnervated muscles were tracked across all recordings to determine whether they were task-specific or shared across multiple tasks. Specifically, within each reinnervated muscle, we assessed whether motor units recruited when the reinnervated muscle acted as an agonist were also active when the same acted as an antagonist. In P1, considering TMR1, 40\% of motor units identified when TMR1 acted as an agonist were also recruited during the corresponding opposite tasks (antagonist role), whereas no shared motor units were identified in TMR2. Similarly, in P3, 60\% and 0\% of recruited motor units were shared between agonist and antagonist tasks for TMR1 and TMR2. On average, considering the three reinnervated muscles of P2 TMR1,2,3, 42.3 $\pm$ 24.7\% of the motor units identified in an agonist reinnervated muscle for a task were also recruited when the same reinnervated muscle was expected to act as antagonist. For example (Fig.~\ref{fig:concept}-e), in P2 six motor units were decomposed from recordings in TMR2 during wrist flexion (on the left, shaded blue), with four of these also active during wrist extension when TMR1 acted as an antagonist (on the right, shaded blue). In Supplementary Fig.\ref{fig:MUAPs} we provide an example of two motor units tracked across agonist and antagonist tasks, when these had an agonist and antagonist role. For each participant and recorded task, the number of motor units recruited during both the agonist and corresponding antagonist tasks, as well as those shared across different tasks, is shown in Fig.~\ref{fig:shared}-a,b,c and highlighted in yellow. For each task and patient, the total number of motor units identified within the respective agonist and antagonist reinnervated muscles is shown in Fig.~\ref{fig:shared}-d (e.g., when P2 performed wrist flexion, 5 and 3 MUs were respectively identified in recordings from the reinnervated agonist (TMR2) and antagonist (TMR3) muscles). As expected, some motor units were also shared across different tasks of the missing limb. Importantly, when considering only agonist tasks for each reinnervated muscle, in P2, while 5 out of the 16 performed tasks did not have independent motor units, the remaining had an average of 92.3 $\pm$ 11\% independent motor units (i.e., not shared across tasks). These results are consistent with previous findings \cite{ferrante2025implanted}. Tasks with no independent motor units had a low number of identified motor units, possibly due to the limited pick-up volume of the micro-electrode array or the participant's difficulty in executing the task. Within each reinnervated muscle, when considering motor units active during agonist tasks and those active when the same reinnervated muscle behaved as antagonist, the percentage of independent motor units decreased to 47.3 $\pm$ 19.9\%, further indicating the task specificity of those motor units always recruited within the same task with an agonist or antagonist role. We do not report these results on task-specificity of motor units in P1 and P3 since the number of recorded tasks as well as the number of motor units identified per task were too low for this specific analysis. For these reasons, the remaining analysis is done considering P2. 
\\
\subsection*{Firing properties of motor units in agonist and antagonist reinnervated muscles}
The mechanism of recruitment and control of motor units for muscle force generation is well documented in individuals without limb deficiency. However, how targeted nerve transfer and biased afferent pathways affect motor unit firing properties remains unclear.
For each task performed by P2, we assessed whether (i) motor units of the same reinnervated muscle, recruited when the muscle had an agonist versus an antagonist role (shared units; e.g.; the firing rate of motor units 1-3 in Fig.~\ref{fig:3}-c recruited for pronation is compared to the firing rate of the same motor units, 1-3 in Fig.~\ref{fig:3}-d, during supination) had similar firing rate properties (null hypothesis); (ii) motor units recruited within the agonist and antagonist reinnervated muscle during a task had similar firing rate properties (e.g., firing rate of motor units in Fig.~\ref{fig:3}-a and Fig.~\ref{fig:3}-c is compared). For each motor unit, the average firing rate was computed as the median (MFR) of the inter-spike intervals (ISI) since these had a non-normal distribution (see Materials and Methods). Moreover, only motor unit spikes occurring during the plateau of the contraction were considered in the MFR computation. The variability of motor unit discharges was quantified using the coefficient of variation, calculated as the ratio between the standard deviation and the median of ISIs. A significant difference (p = 0.002, two-sided Wilcoxon signed rank test) was found between the distribution of the MFR of motor units when these had an agonist or antagonist role (Fig.~\ref{fig:3}-e). In the figure, the two groups, color-coded in blue and light blue, contain the MFR values of motor units recruited when the same reinnervated muscle acted as agonist (blue, Fig.~\ref{fig:3}-d) and antagonist (Fig.~\ref{fig:3}-c) for the executed tasks (i). When individual tasks were examined (ii), no significant difference (p = 0.28, two-sided Wilcoxon signed rank test) was found between MFR of motor units recruited within the agonist and antagonist reinnervated muscles (Fig.~\ref{fig:3}-f, the ``agonist TMR" group includes all motor units recruited within the respective agonist reinnervated muscle, panel a and d; the ``antagonist TMR" group includes motor units in the antagonist muscle). No significant differences were found in recruitment threshold or coefficient of variation. See Materials and Methods for metrics details.

\subsection*{Assessment of coactivation between agonist and antagonist reinnervated muscles}
In individuals with intact limbs, both supraspinal and spinal mechanisms contribute to the coordinated control of agonist and antagonist muscle groups. Depending on task demands and neuromechanical constraints, muscle activation emerges from the projection of common and independent synaptic inputs to the relevant motor neuron populations and integration of afferent inputs. The presence of common input is reflected in shared low-frequency oscillations within motor neurons discharge activity. 
To examine the dimensionality of common synaptic input to motor neurons of reinnervated agonist and antagonist muscles, a neural manifold analysis was performed for each agonist–antagonist task pair. For each pair, we combined the spike trains of motor units recruited within the agonist and antagonist reinnervated muscles during each pair of tasks. For each reinnervated muscle, motor units were tracked across tasks, and the binary spike trains were smoothed using a Hanning window of \SI{0.4}{\second} duration. The total number of unique motor units defined the dimensionality of the motor unit space $V$ (Fig.~\ref{fig:3}-g, left side). We used Non-Negative Matrix Factorization (NNMF) \cite{lee2000algorithms} to estimate the neural manifold $H$ (Fig.~\ref{fig:3}-g, right side) underlying $V$. We estimated the dimension of the neural manifold (i.e., the dimension of $H$) as the minimum number of latent factors beyond which an additional one increased the coefficient of determination $R^{2}$ \cite{muceli2010identifying} between the original data $V$ and the reconstructed data $WH$ by less than 5\% \cite{clark2010merging} (Fig.~\ref{fig:3}-h). We hypothesized the dimensionality of the neural manifold to be at least two, as a result of at least two distinct sources of common input to motor units contributing to the agonist and corresponding opposite (antagonist) tasks. Synchronous patterns of motor unit activation (co-activation) between motor units of the agonist and antagonist reinnervated muscles would imply a manifold dimensionality equal to one, and the inability to distinguish a command for the two opposing tasks.
For all the recorded pairs of agonist-antagonist tasks, the estimated dimensionality was 4. While task complexity as well as instances of non-synchronous motor neuron activity may contribute to increasing the manifold dimensionality, these results show that motor units of the agonist and antagonist reinnervated muscles are not dominated by co-activation despite TMR and perturbed afferent pathways. Additionally, as indicated in the above analysis and shown in fig.\ref{fig:shared}, tasks shared a number of motor units, which may have been assigned to a separate dimensionality depending on their functional role within the task. Repeating the analysis for each task separately (without concatenating agonist and antagonist motor units) confirmed a neural manifold dimensionality greater than 1: across 18 tasks, 43.75\% showed dimensionality 3, and the remainder 4.

\section*{Discussion}
The functional interplay between groups of agonist and antagonist muscles arises from continuous sensory integration into motor commands, shaped by task requirements, constraints, and the physical interaction with the external environment. Under stable conditions, reciprocal inhibition via afferent pathways facilitates agonists contraction by suppressing antagonists motor units \cite{pierrot2005circuitry}. However, co-contraction of agonist and antagonist muscles has been extensively observed in humans when motor tasks are performed under particular requirements (time constraint \cite{poscente2021rapid}, accuracy \cite{danion2004relation}), or when elements of uncertainty may hinder successful task execution \cite{burdet2001central,franklin2003adaptation}. In these cases, tonic coactivation can override spinal reciprocal inhibition, enhancing joint stability and enabling rapid responses to perturbations \cite{nielsen1993regulation}.

Although nerve transfer surgeries (TMR, RPNI) play a critical role in bionic limb reconstruction, they do not restore the native neuro-mechanical coupling between agonist and antagonist muscle groups typical of intact limbs. The loss of this biomechanical interplay profoundly alters afferent signalling, which is essential for movement coordination and execution of the above-mentioned motor control strategies. It therefore remains an open question whether, following amputation and TMR the central nervous system continues to generate and coordinate distinct agonist and antagonist neural commands. This question has remained unresolved, as TMR muscles are typically interfaced using single-channel bipolar surface electrodes that provide only a coarse representation of the neural activity conveyed by polyfascicular nerves. 

For each task performed by the three participants, among the inspected reinnervated muscles, we identified a main agonist and antagonist muscle depending on the role the rerouted nerve had in the intact limb. In P2, where we concurrently record the EMG activity from three reinnervated muscle we can further validate this choice (Fig.~\ref{fig:emgs}). We then provide the first motor-unit–level evidence that functional neural coupling between agonist and antagonist pathways persists after polyfascicular nerve transfers. By combining TMR surgery with high-density intramuscular microelectrode arrays and motor unit decomposition, we isolated and tracked individual motor neurons associated with distinct tasks of the missing limb and examined their functional roles. We found that a substantial fraction of motor units recruited as agonists in one task behaved as antagonists during execution of the opposing task. This demonstrates preserved functional recruitment even when agonist and antagonist neural pathways converge onto the same muscle or are distributed across biomechanically uncoupled muscles. Importantly, motor units recruited when the reinnervated muscle acted as agonist and when it acted as antagonist, exhibited significantly different discharge rates depending on whether they served an agonist or antagonist role. Furthermore, neural manifold analysis revealed a dimensionality greater than one across all agonist–antagonist task pairs, indicating that their activity was not dominated by synchronous activation.

Although this study did not directly assess voluntary modulation of both agonist and antagonist commands, and therefore cannot determine whether the observed patterns reflect a strategic adaptation (e.g., to maintain tracking precision) or altered excitation–inhibition balance within reorganized circuits, functional specific recruitment of motor units in agonist and antagonist reinnervated muscles provides evidence of persistent residual neural coupling between agonist and antagonist pathways, despite anatomical reorganisation following amputation and targeted nerve transfer. In turn, this suggests that such neural coupling is at least partially mediated by spinal and supraspinal mechanisms that remain intact post-amputation.

In individuals with limb loss, the presence of persistent functional neural coupling between agonist and antagonist commands, even after targeted nerve transfer, may support the rationale behind surgeries like the Agonist-Antagonist Myoneural Interface (AMI), which aim to restore mechanical and thus neural coupling between reinnervated muscles to promote natural proprioception \cite{clites2018proprioception} and their surgical implementation even in TMR patients. Indeed, the integration of sensory feedback may allow participants to further tune the coordination of motor units recruited within agonist and antagonist reinnervated muscles and could provide deeper insight into the mechanisms underlying our findings. Moreover, such persistent functional neural coupling may be exploited for controlling advanced prostheses with variable stiffness actuators (which require modulation of agonist and antagonist commands), allowing amputees to modulate the dynamics of movement besides the kinematics, and for artificial substitution of sensory feedback associated with joint impedance and movement perception.
Finally, these results further support TMR of polyfascicular nerves, demonstrating the high information transfer capacity of this biological interface when reinnervated muscles are coupled with high-resolution electromyography systems capable of decoding motor intent at the level of individual motor unit recruitment. The ability to detect such complex neural signals (agonist, antagonist, synergist signals) may enable the development of advanced neural interfaces and prosthetic control systems that more closely replicate natural human motor behaviour.
\section*{Materials and Methods}\label{sec:methods}
\subsection*{Participant}
The current study was approved by the Ethics Committee of Imperial College (reference number: 19IC5641) and performed according to the Declaration of Helsinki. Three participants (males; 62, 65, 53 years old) who suffered from an amputation of their upper-limb and underwent TMR surgery were recruited for this study. Each participant provided written informed consent before taking part in the experimental session. 
P1 suffered an accident in 1994 that resulted in a brachial plexus injury and subsequent left transhumeral amputation. Twenty years after the injury, TMR surgery was performed, resulting in muscle signals in his deltoid muscle, now being mediated by the ulnar and median nerves in the anterior part and by the radial nerve in the posterior part of the muscle. Control of the myoelectric prosthesis was achieved using two bipolar surface electrodes.
P2 underwent transhumeral amputation of his left upper limb and TMR surgery by nerve-to-nerve coaptation (\cite{pettersen2024targeted}), mainly to treat phantom limb pain, 8 years before this study. The ulnar nerve was transferred to the pectoralis minor muscle, the median nerve to the lower (abdominal) part of the pectoralis major muscle, and the radial nerve to the latissimus dorsi muscle. The pectoralis minor was released from the coracoid process and was transferred into the axilla to enable easier detection of the EMG signals. The participant daily uses a myoelectrical prosthesis, controlled with two standard surface bipolar electrodes positioned on the remaining (non-reinnervated) triceps and biceps muscles of the limb. Twenty years before this study, P3 had his left arm amputated at the glenohumeral level and underwent TMR surgery via nerve-nerve coaptation \cite{pettersen2024targeted} 8 years after amputation. The musculocutaneous nerve was transferred to the clavicular part of the pectoralis major, the ulnar nerve to the pectoralis minor, the median nerve to the sternocostal and abdominal part of the pectoralis major and the radial nerve to the latissimus dorsi and teres major. After this procedure, he could use 6 surface bipolar electrodes in order to operate his myoelectric prosthesis.
Patient details are summarised in Table~\ref{tab:patient}.
\begin{table*}[h!]
\centering
\footnotesize
\setlength{\tabcolsep}{3pt}
\renewcommand{\arraystretch}{1.05}

\newcolumntype{L}{>{\raggedright\arraybackslash}p{0.17\linewidth}}
\newcolumntype{Y}{>{\raggedright\arraybackslash}X}

\caption{Patient characteristics, surgery details, prosthetic information, and experimental protocol. h: hours; MVC: maximal voluntary contraction; TMR: targeted muscle reinnervation.}
\label{tab:patient}
\begin{tabularx}{\linewidth}{L Y Y Y}
\toprule
\textbf{TMR volunteers} & \textbf{P1} & \textbf{P2} & \textbf{P3} \\
\midrule
\multicolumn{4}{l}{\textbf{Patient characteristics}} \\
Age & 62 y & 63 y & 53 y \\
Time since amputation & 28 y & 45 y & 20 y \\
Level of amputation & transhumeral & transhumeral & glenohumeral \\
\midrule
\multicolumn{4}{l}{\textbf{Surgery details}} \\
Time from TMR surgery & 8 y & 8 y & 13 y \\
TMR site and innervating nerve
& Anterior deltoid: ulnar + median (TMR1); posterior deltoid: radial (TMR2)
& Pec.\ major (abdom.): median (TMR1); pec.\ minor: ulnar (TMR2); latissimus dorsi: radial (TMR3)
& Pec.\ major (clav.): MC; pec.\ minor: ulnar (TMR1); pec.\ major (sternocost.): median I; pec.\ major (abdom.): median II; latissimus dorsi/teres major: radial (TMR2) \\
\midrule
\multicolumn{4}{l}{\textbf{Prosthesis information}} \\
Prosthetic type & Myoelectric & Myoelectric & Myoelectric \\
Prosthetic use & 16 h/day & 15 h/day & 0--12 h/day, 1--2 d/week \\
No.\ EMG sensors & 2 & 2 & 6 \\
Movements
& Fist; elbow extension
& Elbow flexion; extension
& Tripod; finger ext.; elbow flex.--ext.; pron.--sup. \\
Natural phantom position
& 30° elbow flex.; forearm neutral; hand slightly closed
& 20° elbow flex.; wrist neutral; 30° abd.; hand slightly open
& 90° elbow flex.; forearm neutral; hand closed \\
\midrule
\multicolumn{4}{l}{\textbf{Protocol}} \\
Target contraction
& Ramps 4 s; iso 10 s at 10\% MVC
& Ramps 4 s; iso 10 s at 10\% MVC
& Ramps 2 s; iso 5 s at 10\% MVC \\
\bottomrule
\end{tabularx}
\end{table*}

\subsection*{Apparatus for intramuscular EMG recordings in reinnervated muscles}
A microelectrode array, described by Muceli {\it{et al.}} \cite{muceli2022blind}, was used to probe each reinnervated muscle. The microelectrode array consists of a double-sided polyimide structure \SI{20}{\micro\metre} thick, with 40 platinum channels (area of \SI{5357}{\micro\metre\squared}) linearly distributed over \SI{2}{\cm} with an interelectrode distance of \SI{500}{\micro\metre}. For each reinnervated muscle, the region exhibiting the highest myoelectric activity was identified through clinical examination (including palpation and visual assessment of muscle contractions during tasks involving movements of the missing limb associated with the transferred nerve) surface EMG measurements using the MyoBoy system (Ottobock Healthcare Products GmbH, Duderstadt, Germany). This region was selected as the insertion site for the micro-array. Following skin disinfection, the micro-arrays were acutely inserted into the muscle at a shallow angle using a hypodermic needle comparable in size to those employed in conventional concentric needle recordings. The needle was then removed while ensuring that the microelectrode array remained securely positioned within the muscle. The entire insertion procedure was performed using a portable ultrasound scanner. At the end of the experimental session, the microelectrode array was removed and the skin was disinfected.

The EMG signals were recorded in monopolar configuration using a multichannel amplifier (Quattrocento, OT Bioelettronica, Torino, Italy) with a gain of 150 and band-pass-filtered (10-\SI{4400}{\hertz}) before being sampled at \SI{10240}{\hertz} using an A/D converter to \SI{16}{\bit}. The reference and ground electrodes were placed in areas of no significant myoelectric activity on the residual limb or bony parts of the shoulder. 

Two microelectrode arrays were inserted in P1 and P3, while 3 reinnervated muscles were probed in P2.
\subsection*{Experimental protocol}
Each participant was instructed to perform predefined movement of their missing limb, which led to contraction of reinnervated muscles with inserted microelectrode arrays. The list of movement included in the protocol was tailored on participant-specific functional capabilities (e.g., ability to sustain prolonged contractions and ability to execute the task) and determined according to the nerve transfer matrix. Additionally, for each task a reinnervated muscles was defined as agonist based on the original functional role the rerouted nerve ad prior to amputation. 

Each participant was seated in front of a computer display providing real-time visual feedback of EMG activity recorded from the pre-defined agonist reinnervated muscle via the intramuscular micro-electrode array. Feedback was derived from the bipolar channel exhibiting the maximum signal amplitude. For each task, the maximum voluntary contraction (MVC) was acquired at the beginning of the trial and used to normalize the corresponding EMG signals. During task execution, the participant was then required to modulate muscle activation to track a predefined trapezoidal target profile displayed on the screen. A trapezoidal profile consisted of (i) an ascending ramp phase, during which EMG amplitude had to be increased to a specified percentage of MVC; (ii) a plateau phase, during which EMG amplitude had to be maintained at the targeted percentage of MVC; and (iii) a descending ramp phase, where EMG amplitude had to be reduced until complete muscle relaxation was achieved. Twenty seconds of rest was allocated between each contraction to minimize fatigue with longer rest when needed.
\subsection*{Intramuscular EMG signals processing}
The recorded intramuscular EMG signals were high-pass filtered at \SI{1000}{\hertz} with a zero-lag first-order digital filter. The quality of the signals was assessed by computing the root-mean-square of \SI{5}{\second} of data recorded at rest before starting the task recording. Channels yielding a baseline noise $>$ \SI{15}{\micro\volt} were visually inspected and removed.
\subsection*{Motor unit decomposition}\label{sec4:sec3_decom}
Preprocessed EMG signals were decomposed into their constituent motor unit spike trains using the algorithm described by Holobar and Zazula \cite{holobar2007multichannel} and validated by Muceli {\it{et al.}} \cite{muceli2022blind} using the same micro-arrays adopted in this study. Data from each task were individually decomposed. The decomposition was subsequently validated by an expert investigator using EMGLAB \cite{mcgill2005emglab}. Each EMG signal was systematically inspected to identify decomposition errors, including missed discharges and incorrectly assigned spikes, with particular attention to abnormally long or short interspike intervals. Detection and resolution of superimposed motor unit action potentials (MUAPs) were facilitated by the use of multi-channel intramuscular recordings, as individual channels of the microelectrode array sampled distinct regions of a motor unit’s territory and therefore provide complementary observations of its electrical activity. Although MUAP amplitude and waveform shape varies across channels, the underlying motor unit firing pattern remains consistent. This spatially redundant information was exploited to disambiguate superimposed action potentials originating from different motor units.

After complete identification of a motor units spike train, the corresponding MUAP template was subtracted from the EMG signals using EMGLAB. Decomposition was deemed complete when the residual signal power was comparable to the baseline noise level. Low-amplitude potentials and signals exhibiting burst-like firing patterns were excluded due to insufficient decomposition reliability. Duplicates spike trains due to decomposition errors or presence of satellite potentials were removed. This procedure was repeated for each microelectrode array channel. 
Motor units exhibiting sparse discharge activity were excluded from the analysis. Only motor units with spike trains identified at a Pulse-to-Noise Ratio (PNR; a signal-to-interference metric that quantifies the reliability with which a motor unit’s activity can be distinguished from background noise in the EMG signal) greater than 30 dB were retained \cite{holobar2014accurate}. 
\subsection*{Motor units firing properties}\label{sec4:sec4_firing}
The following analysis was performed for each motor unit recruited for a task. For each motor motor unit spike train, inter-spike intervals longer than \SI{250}{\milli\second} were considered as reflecting pauses in motor unit tonic activity and were removed from the inter-spike intervals of the motor unit. The distribution of the inter-spike intervals was tested for normality using the D'Agostino-Pearson's test ($\alpha$ = 0.95) \cite{d1990suggestion} before computing the motor unit average firing rate. The average firing rate was defined as the median firing rate (MFR), computed as the median of the inter-spike intervals of motor units since these had a non-normal distribution. The variability of the motor unit discharges was quantified using the Coefficient of Variation CoV [\%], computed as the ratio between the median and the standard deviation of ISIs. The istantanous motor unit firing rate was obtained by applying an Hanning window of \SI{0.4}{\second} over the impulse train corresponding to the firing times of that motor unit.
\subsection*{Motor unit tracking across tasks}
\label{sec:method:mu_tracking}
For each reinnervated muscle, motor units were tracked across multiple tasks to determine whether they were recruited across multiple tasks (shared) or selectively recruited during a single task (task-specific). For each pair of motor units of two tasks we calculated the (i) coefficient of determination between the normalised Motor Unit Action Potential (MUAPs) template obtained by spike-triggered averaging in a 20 ms window on the channel where the peak-to-peak unipolar amplitude of MUAPs was the largest; (ii) the distribution of the MUAP across channels was estimated by concatenating the average MUAPs obtained for each channel to build an image of the spatial distribution of the MUAP potential (channels x time x motor unit amplitude). A 15\% threshold was applied to the absolute amplitude of pixels and used to segment the area occupied by the MUAP. The MUAP cross-section (i.e., channels spanned by MUAP) was obtained by projecting the segmented area on the channel axis. Two MUAPs were flagged as belonging to the same motor units if $\rho \geq 0.85$ or if their cross-section overlapped; visual inspection assessed the match between two motor units.
\subsection*{Neural manifold estimation}\label{sec:nmf}
We use Non Negative Matrix Factorization (NNMF) to estimate the neural manifold $\textbf{H}$ embedded in the motor unit space $\textbf{V}$ defined by (i) concatenating instantaneous firing rate of motor units recruited during a single task and recorded from the agonist and antagonist reinnervated muscles and (ii) by concatenating instantaneous firing rate of motor units recorded during the agonist and corresponding antagonist task, recorded from respective agonist and antagonist reinnervated muscles. For each task or task pair, the matrix $\textbf{V}$, of dimension $m \times t$, is obtained by (i) or (ii), and by normalising the smoothed spike-trains to have unit variance. The variable $m$ indicates the total number of independent motor units, shared motor units have their istantanous firing rate located in the same row of $\textbf{V}^{i}$. According to NNMF, $\textbf{V}$ is mathematically described as follows:
\begin{equation}
\label{eq:model}
\textbf{V} \approx \textbf{W}\textbf{H}
\end{equation}
where $\textbf{W}$ is the non-negative basis matrix $\textbf{V}$ of dimension $m$ by $l$, and $\textbf{H}$ of dimension $l$ by ($r \times t$) is the non-negative matrix of latent time-dependent variables. Each column of $\textbf{W}$ represents the contribution of the m motor units to the latent signals $l$. Each row of $\textbf{H}$ is a latent variable comprising the time-dependent input to motor units.
NNMF solves a non-convex optimisation problem (i.e., minimises the reconstruction error defined as the Euclidean distance between $\textbf{V}$ and $\textbf{W}\textbf{H}$), prone to converging to local minima. For this reason, given $l$, NNMF is repeated 10 times with random initialisation of $\textbf{W}$ and $\textbf{H}$ \cite{cheung2009stability} and random shuffling of motor units spike trains in $V$ and task repetitions concatenation order; the maximum number of iterations is set to 100. The latent space dimension $l$ is set equal to the minimum number of latent factors explaining a significant portion of the total variance within $\textbf{V}^{i}$. In our study, NNMF is applied for values of $l$ ranging from 1 to 5. The coefficient of determination $R^{2}$ is computed as in Muceli {\it{et al.}} \cite{muceli2010identifying}. The MSE values are reported for completeness. The number of latent factors is determined using the $R^{2}$-curve as in d'Avella {\it{et al.}} \cite{d2003combinations} and Clark {\it{et al.}} \cite{clark2010merging}. The above procedure was also followed in the study the current work builds upon \cite{ferrante2025implanted}.
\subsection*{Statistical analysis and reproducibility}\label{sec4:sec7_statistic}
Shapiro–Wilk tests were used to confirm homogeneity of variance and normal distribution of data, respectively. 
Kruskal–Wallis one-way Anova analysis was undertaken since data violated parametric assumptions and used to analyse differences between motor units median firing rates.  The D’Agostino-Pearson’s test was used to test the inter-spike intervals distribution for normality. Statistical significance was assumed at P $\leq$ 0.05. All data is reported as mean $\pm$ standard deviation. 

The following procedures were implemented to ensure reproducibility: (i) only motor units that could be reliably decomposed were included in the analysis; (ii) EMG signals were processed using an automatic decomposition algorithm extensively validated across multiple datasets, with results manually reviewed using the spike-sorting software EMGLAB to identify and correct decomposition errors; and (iii) motor unit tracking was performed automatically and subsequently visually inspected by an experienced examiner to verify accuracy.

\clearpage 

%
\bibliography{science_template} 

@article{mcgill2005emglab,
  title={EMGLAB: an interactive EMG decomposition program},
  author={McGill, Kevin C and Lateva, Zoia C and Marateb, Hamid R},
  journal={Journal of neuroscience methods},
  volume={149},
  number={2},
  pages={121--133},
  year={2005},
  publisher={Elsevier}
}

@article{muceli2022blind,
  title={Blind identification of the spinal cord output in humans with high-density electrode arrays implanted in muscles},
  author={Muceli, Silvia and Poppendieck, Wigand and Holobar, Ale{\v{s}} and Gandevia, Simon and Liebetanz, David and Farina, Dario},
  journal={Science advances},
  volume={8},
  number={46},
  pages={eabo5040},
  year={2022},
  publisher={American Association for the Advancement of Science}
}

@article{holobar2014accurate,
  title={Accurate identification of motor unit discharge patterns from high-density surface EMG and validation with a novel signal-based performance metric},
  author={Holobar, Ale{\v{s}} and Minetto, Marco Alessandro and Farina, Dario},
  journal={Journal of neural engineering},
  volume={11},
  number={1},
  pages={016008},
  year={2014},
  publisher={IOP Publishing}
}

@article{holobar2007multichannel,
  title={Multichannel blind source separation using convolution kernel compensation},
  author={Holobar, Ales and Zazula, Damjan},
  journal={IEEE Transactions on Signal Processing},
  volume={55},
  number={9},
  pages={4487--4496},
  year={2007},
  publisher={IEEE}
}

@article{d1990suggestion,
  title={A suggestion for using powerful and informative tests of normality},
  author={D'agostino, Ralph B and Belanger, Albert and D'Agostino Jr, Ralph B},
  journal={The American Statistician},
  volume={44},
  number={4},
  pages={316--321},
  year={1990},
  publisher={Taylor \& Francis}
}

@article{kuiken2004use,
  title={The use of targeted muscle reinnervation for improved myoelectric prosthesis control in a bilateral shoulder disarticulation amputee},
  author={Kuiken, Todd A and Dumanian, Gregory Ara and Lipschutz, Robert D and Miller, Laura A and Stubblefield, KA},
  journal={Prosthetics and orthotics international},
  volume={28},
  number={3},
  pages={245--253},
  year={2004},
  publisher={Taylor \& Francis}
}

@article{pettersen2024targeted,
  title={Targeted Muscle Reinnervation: Surgical Protocol for a Randomized Controlled Trial in Postamputation Pain},
  author={Pettersen, Emily and Sassu, Paolo and Pedrini, Francesca Alice and Granberg, Hannes and Reinholdt, Carina and Breyer, Juan Manuel and Roche, Aidan and Hart, Andrew and Ladak, Adil and Power, Hollie A and others},
  journal={Journal of Visualized Experiments},
  volume={2024},
  number={205},
  pages={e66379},
  year={2024},
  publisher={MYJoVE Corporation}
}

@article{lee2000algorithms,
  title={Algorithms for non-negative matrix factorization},
  author={Lee, Daniel and Seung, H Sebastian},
  journal={Advances in neural information processing systems},
  volume={13},
  year={2000}
}

@article{muceli2010identifying,
  title={Identifying representative synergy matrices for describing muscular activation patterns during multidirectional reaching in the horizontal plane},
  author={Muceli, Silvia and Boye, Andreas Tr{\o}llund and d'Avella, Andrea and Farina, Dario},
  journal={Journal of neurophysiology},
  volume={103},
  number={3},
  pages={1532--1542},
  year={2010},
  publisher={American Physiological Society Bethesda, MD}
}

@article{cheung2009stability,
  title={Stability of muscle synergies for voluntary actions after cortical stroke in humans},
  author={Cheung, Vincent CK and Piron, Lamberto and Agostini, Michela and Silvoni, Stefano and Turolla, Andrea and Bizzi, Emilio},
  journal={Proceedings of the National Academy of Sciences},
  volume={106},
  number={46},
  pages={19563--19568},
  year={2009},
  publisher={National Acad Sciences}
}

@article{clark2010merging,
  title={Merging of healthy motor modules predicts reduced locomotor performance and muscle coordination complexity post-stroke},
  author={Clark, David J and Ting, Lena H and Zajac, Felix E and Neptune, Richard R and Kautz, Steven A},
  journal={Journal of neurophysiology},
  volume={103},
  number={2},
  pages={844--857},
  year={2010},
  publisher={American Physiological Society Bethesda, MD}
}

@article{d2003combinations,
  title={Combinations of muscle synergies in the construction of a natural motor behavior},
  author={d'Avella, Andrea and Saltiel, Philippe and Bizzi, Emilio},
  journal={Nature neuroscience},
  volume={6},
  number={3},
  pages={300--308},
  year={2003},
  publisher={Nature Publishing Group US New York}
}

@article{ferrante2025implanted,
  title={Implanted microelectrode arrays in reinnervated muscles allow separation of neural drives from transferred polyfunctional nerves},
  author={Ferrante, Laura and Boesendorfer, Anna and Barsakcioglu, Deren Y and Baumgartner, Benedikt and Al-Ajam, Yazan and Woollard, Alexander and Kang, Norbert V and Aszmann, Oskar C and Farina, Dario},
  journal={Nature Biomedical Engineering},
  pages={1--16},
  year={2025},
  publisher={Nature Publishing Group UK London}
}

@article{kung2014regenerative,
  title={Regenerative peripheral nerve interface viability and signal transduction with an implanted electrode},
  author={Kung, Theodore A and Langhals, Nicholas B and Martin, David C and Johnson, Philip J and Cederna, Paul S and Urbanchek, Melanie G},
  journal={Plastic and reconstructive surgery},
  volume={133},
  number={6},
  pages={1380--1394},
  year={2014},
  publisher={LWW}
}

@article{proske2012proprioceptive,
  title={The proprioceptive senses: their roles in signaling body shape, body position and movement, and muscle force},
  author={Proske, Uwe and Gandevia, Simon C},
  journal={Physiological reviews},
  year={2012},
  publisher={American Physiological Society Bethesda, MD}
}

@book{pierrot2005circuitry,
  title={The circuitry of the human spinal cord: its role in motor control and movement disorders},
  author={Pierrot-Deseilligny, Emmanuel and Burke, David},
  year={2005},
  publisher={Cambridge university press}
}

@article{burdet2001central,
  title={The central nervous system stabilizes unstable dynamics by learning optimal impedance},
  author={Burdet, Etienne and Osu, Rieko and Franklin, David W and Milner, Theodore E and Kawato, Mitsuo},
  journal={Nature},
  volume={414},
  number={6862},
  pages={446--449},
  year={2001},
  publisher={Nature Publishing Group UK London}
}

@article{danion2004relation,
  title={The relation between force magnitude, force steadiness, and muscle co-contraction in the thumb during precision grip},
  author={Danion, Fr{\'e}d{\'e}ric and Gall{\'e}a, C{\'e}cile},
  journal={Neuroscience letters},
  volume={368},
  number={2},
  pages={176--180},
  year={2004},
  publisher={Elsevier}
}

@article{poscente2021rapid,
  title={Rapid feedback responses parallel the urgency of voluntary reaching movements},
  author={Poscente, Sophia V and Peters, Ryan M and Cashaback, Joshua GA and Cluff, Tyler},
  journal={Neuroscience},
  volume={475},
  pages={163--184},
  year={2021},
  publisher={Elsevier}
}

@article{franklin2003adaptation,
  title={Adaptation to stable and unstable dynamics achieved by combined impedance control and inverse dynamics model},
  author={Franklin, David W and Osu, Rieko and Burdet, Etienne and Kawato, Mitsuo and Milner, Theodore E},
  journal={Journal of neurophysiology},
  year={2003},
  publisher={American Physiological Society}
}

@article{huang2018voluntary,
  title={Voluntary control of residual antagonistic muscles in transtibial amputees: reciprocal activation, coactivation, and implications for direct neural control of powered lower limb prostheses},
  author={Huang, Stephanie and Huang, He},
  journal={IEEE Transactions on Neural Systems and Rehabilitation Engineering},
  volume={27},
  number={1},
  pages={85--95},
  year={2018},
  publisher={IEEE}
}

@article{zuniga2018coactivation,
  title={Coactivation index of children with congenital upper limb reduction deficiencies before and after using a wrist-driven 3D printed partial hand prosthesis},
  author={Zuniga, Jorge M and Dimitrios, Katsavelis and Peck, Jean L and Srivastava, Rakesh and Pierce, James E and Dudley, Drew R and Salazar, David A and Young, Keaton J and Knarr, Brian A},
  journal={Journal of neuroengineering and rehabilitation},
  volume={15},
  number={1},
  pages={48},
  year={2018},
  publisher={Springer}
}

@article{clites2018proprioception,
  title={Proprioception from a neurally controlled lower-extremity prosthesis},
  author={Clites, Tyler R and Carty, Matthew J and Ullauri, Jessica B and Carney, Matthew E and Mooney, Luke M and Duval, Jean-Fran{\c{c}}ois and Srinivasan, Shriya S and Herr, Hugh M},
  journal={Science Translational Medicine},
  volume={10},
  number={443},
  pages={eaap8373},
  year={2018},
  publisher={American Association for the Advancement of Science}
}

@article{festin2024creation,
  title={Creation of a biological sensorimotor interface for bionic reconstruction},
  author={Festin, Christopher and Ortmayr, Joachim and Maierhofer, Udo and Tereshenko, Vlad and Blumer, Roland and Schmoll, Martin and Carrero-Rojas, G{\'e}nova and Luft, Matthias and Laengle, Gregor and Farina, Dario and others},
  journal={Nature Communications},
  volume={15},
  number={1},
  pages={5337},
  year={2024},
  publisher={Nature Publishing Group UK London}
}

@article{farina2026decoding,
  title={Decoding Spikes From Multiunit Data},
  author={Farina, Dario and Yu, Tianyi},
  journal={IEEE Reviews in Biomedical Engineering},
  year={2026},
  publisher={IEEE}
}

@article{nielsen1993regulation,
  title={The regulation of presynaptic inhibition during co-contraction of antagonistic muscles in man.},
  author={Nielsen, J and Kagamihara, Y},
  journal={The Journal of physiology},
  volume={464},
  number={1},
  pages={575--593},
  year={1993},
  publisher={Wiley Online Library}
}
\bibliographystyle{sciencemag}

%
%
%
%
%
%


\section*{Acknowledgments}
The authors thank Aritra Kundu for contributing to the manual editing of motor units and Aron Cserveny
(https://www.sciencevisual.at/) for conveying key concepts of the work through exceptional illustrations.
\paragraph*{Funding:}
This work was supported by the European Research Council Synergy Grant Natural BionicS (810346).
\paragraph*{Author contributions:}
D.F, L.F., O.C.A. and A.B. conceived this study; L.F., A.B., B.B. and O.C.A. performed the data acquisition; L.F. conducted the analysis; L.F. and D.F. interpreted the data; L.F. contributed to the first draft of the manuscript; all authors edited the manuscript for important scientific content, and all approved the final version.
\paragraph*{Competing interests:}
There are no competing interests to declare.
\paragraph*{Data and materials availability:}
All data supporting this study are provided within the paper. Raw data are available from the corresponding authors upon reasonable request.

\subsection*{Supplementary materials}
\begin{figure}[h]
\centering
\includegraphics[width=1\textwidth]{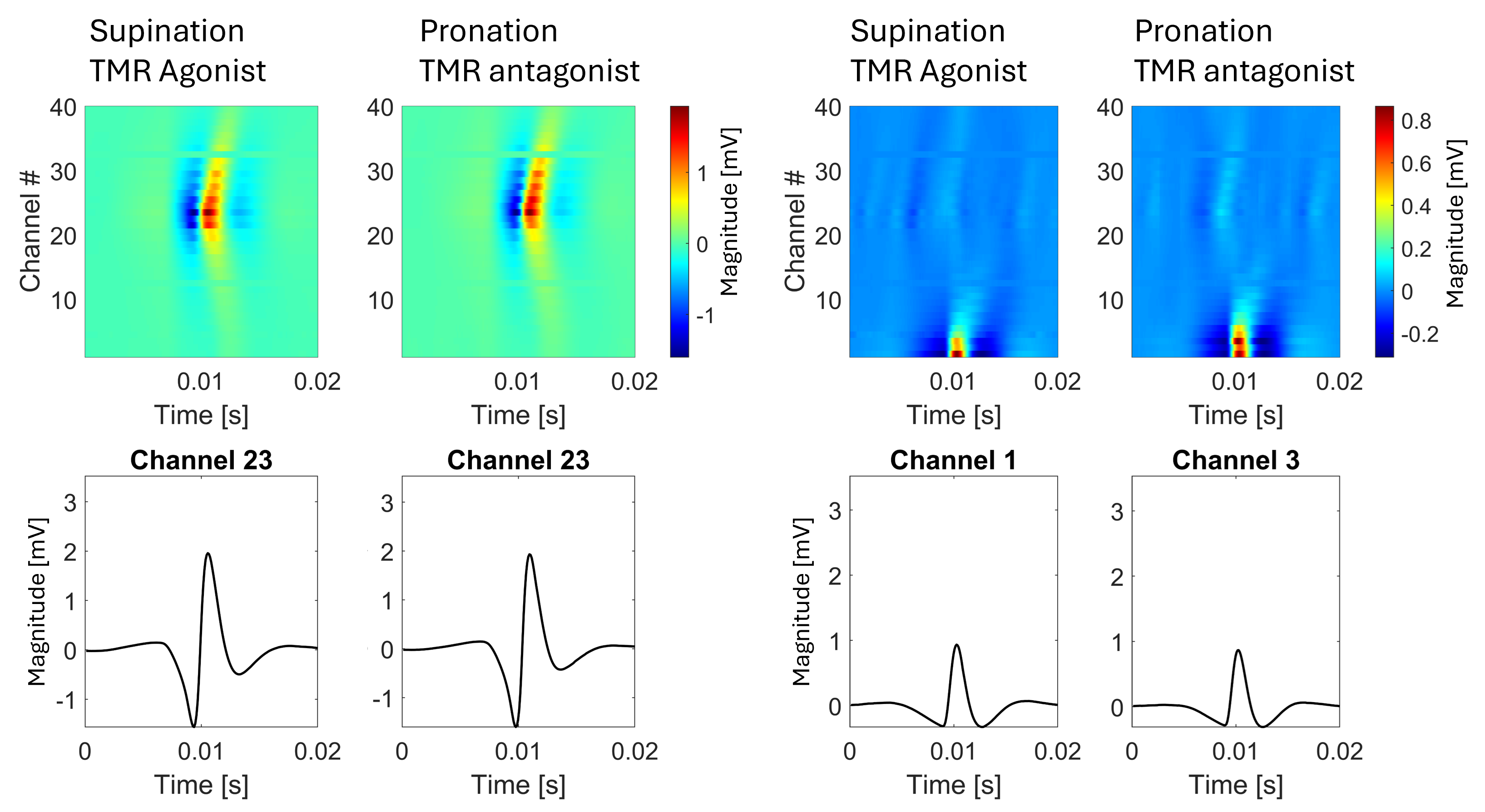}
\caption{\textbf{Supplementary Material. Motor unit tracking across agonist and antagonist tasks}. Distribution of average motor unit action potential across the channels of the microelectrode array inserted in TMR3 of P2 (radial nerve), obtained by spike-trigger averaging. We show two examples of motor unit tracking. On the left, the same motor unit is recruited and detected during supination when the TMR muscle acts as agonist, and during Pronation, when the same muscle acts as antagonist. Similarly, another motor unit is detected in the two conditions, and shown on the right.}
\label{fig:MUAPs}
\end{figure}

\begin{figure}
\centering
\includegraphics[width=0.8\textwidth]{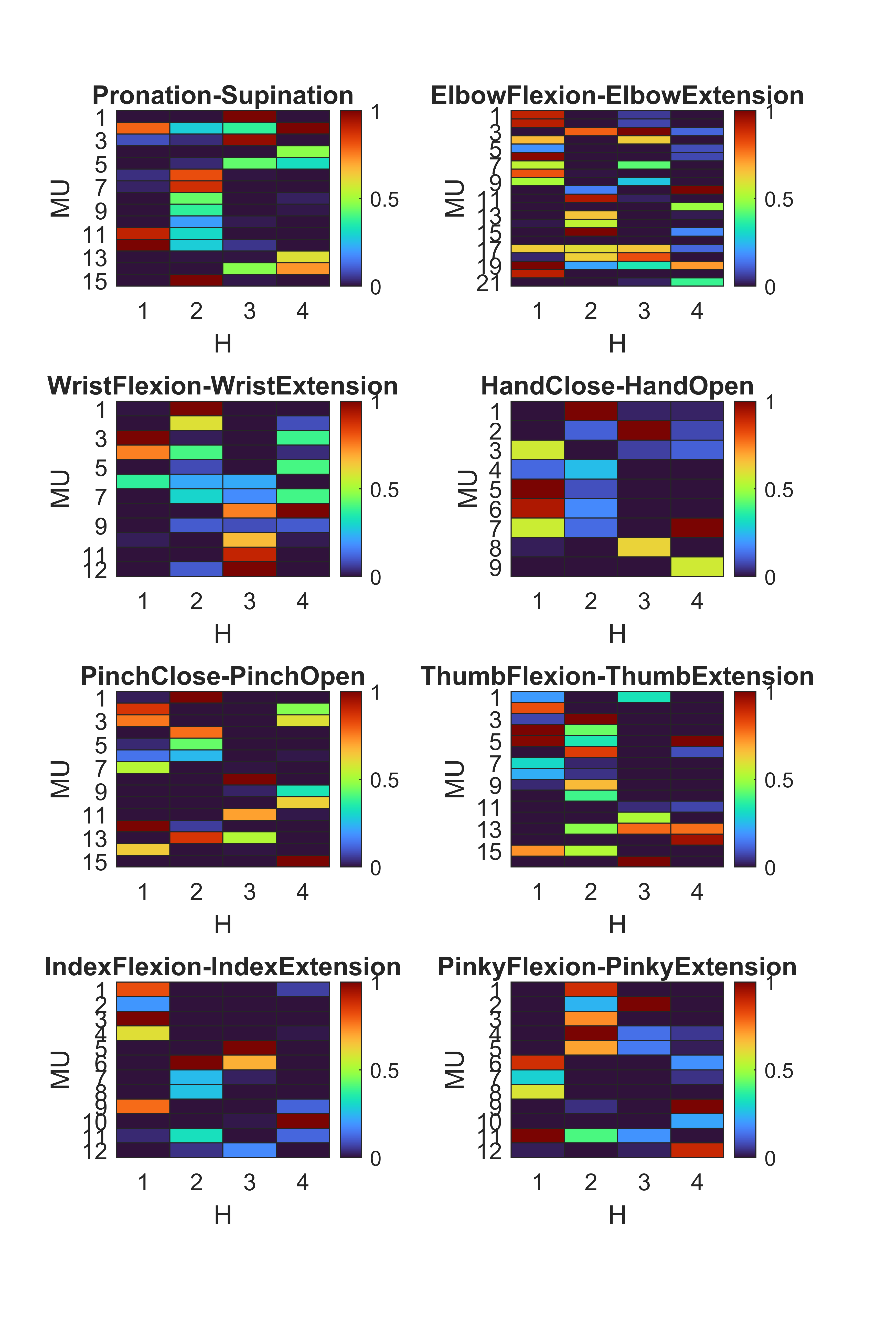}
\caption{\textbf{Supplementary Material. Neural manifold analysis results.} Non-negative matrix of normalized contributions (W) of MUs to latent components output by NNMF for all task pairs.}
\label{fig:W12}
\end{figure}

\begin{figure}
\centering
\includegraphics[width=1\textwidth]{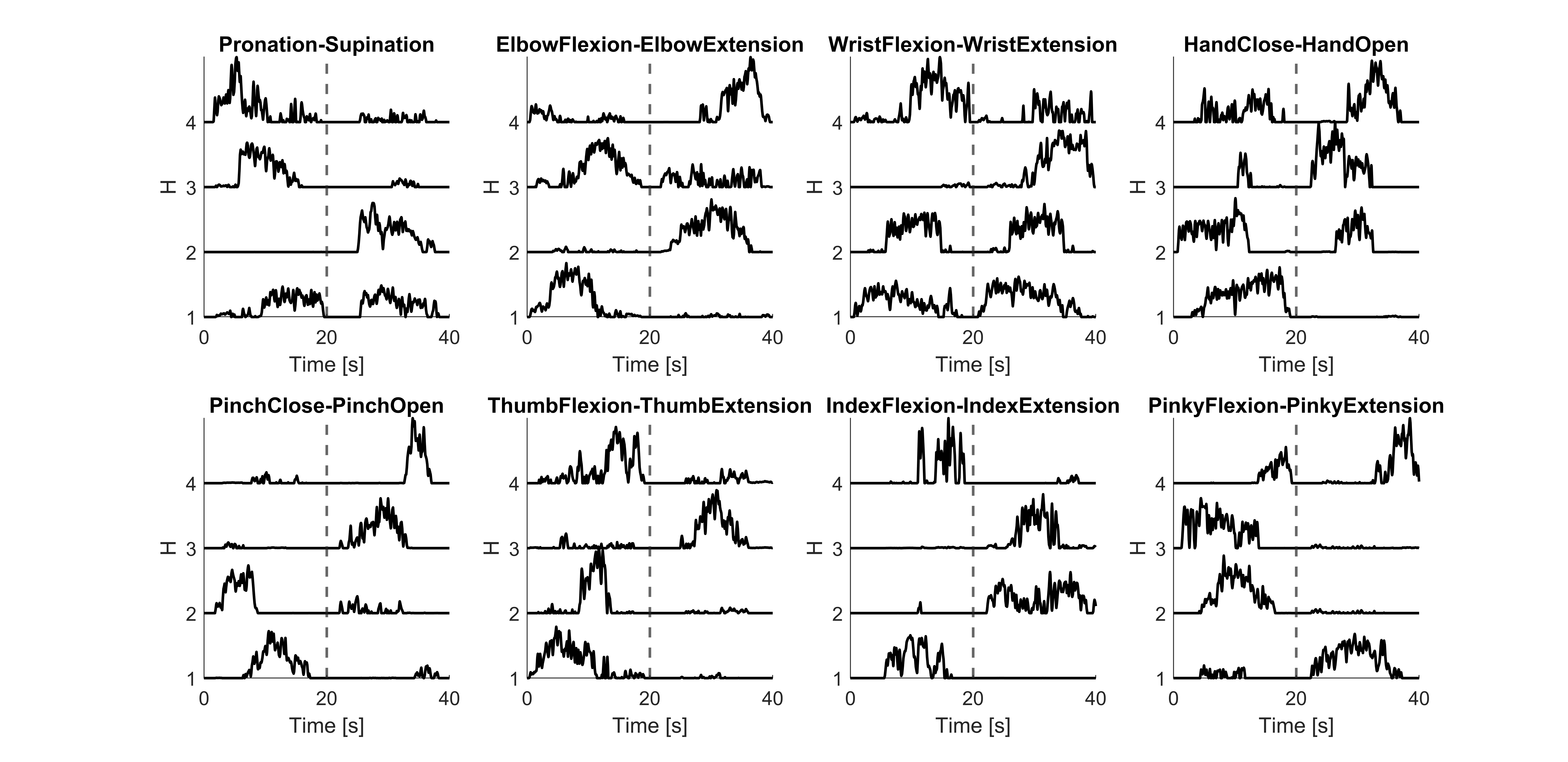}
\caption{\textbf{Supplementary Material. Neural manifold analysis results.} The time-varying latent signals H are shown for each task pair. The estimated dimensionality was 4 for all task pairs.}
\label{fig:H12}
\end{figure}

Two examples of motor unit tracking across an agonist and corresponding antagonist tasks are shown in Fig.~\ref{fig:MUAPs}. Neural manifold analysis results are shown for all pairs of tasks in P2 (Fig.\ref{fig:H12}, Fig.\ref{fig:W12}).

\end{document}